# Statistical analysis of narrow-band signals at setilive.org

Igor Nikitin
Department of High Performance Analytics
Fraunhofer Institute for Algorithms and Scientific Computing
Schloss Birlinghoven, 53757 Sankt Augustin, Germany

igor.nikitin@scai.fraunhofer.de

**Abstract.** SETILive is a web project forwarding radio signals from SETI Institute's Allen Telescope Array (ATA) for the analysis of volunteers. It contains a large archive with more than 1.5 millions observations for more than 7.5 thousands observation targets, including directions to exoplanets discovered by telescope Kepler and other sources. It also supports various tools for signal collection and classification. Till recent time it supported live feeds of signals from ATA together with a feedback loop, a possibility to interrupt the schedule and repeat the observation of an interesting signal registered by sufficiently many viewers. Unfortunately, since 12-Oct-2014 the live feeds have been discontinued. We hope that the project will persist, taking into account the importance of the search subject, the worldwide interest to the topic and the value of already collected data. In this paper we present the results of statistical analysis of data stored in SETILive archive, using Radon transform and specially constructed filter for selection of single beams, potential signals of ET origination. We will also estimate statistical significance of signals depending on their signal-to-noise ratio using Monte Carlo simulation and select 28 strong signals and totally 1072 statistically significant signals in the archive.

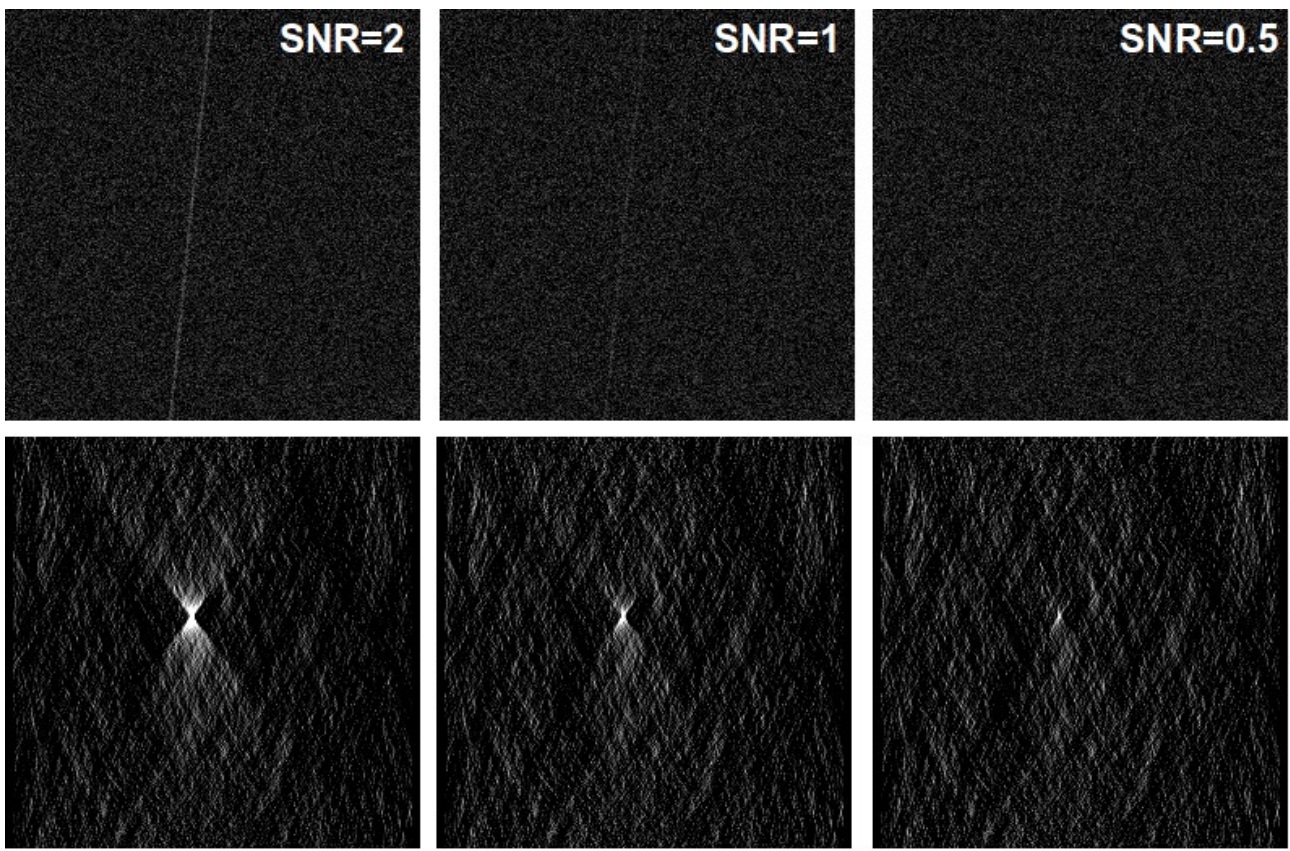

**Fig.1.** Synthetic signal of different strength and its Radon transform.

## Introduction

Signal analysis at SETI [1] uses so called waterfall plots, representing a moving window Fourier decomposition of the signal with frequency taken as horizontal axis and time as vertical axis. Especially interesting features on the waterfall plots are straight lines, corresponding to narrow signals with slowly drifting frequency. Given a source signal of a fixed frequency, Doppler effect will produce a frequency shift, while the rotation and orbital motion of the Earth and similar motions of the signal source will make this shift varying in time. If the observation period is much less than the periods of these motions, the drift will be approximately linear in time and will produce straight lines on the waterfall plots. Earlier it has been proposed to use Radon transform for searching straight lines on the waterfall plots [2,3]. Radon transform performs an integration of a signal along straight lines in various directions, accumulating lines to points. In this way signal-to-noise ratio (SNR) can be increased. An example on Fig.1 shows a synthetic signal of the form

$$s(f,t) = SNR \exp(-((f-f_0-kt)/b)^2) + N(f,t)$$

where f denotes frequency, t – time, $f_0+kt$ represents a line on waterfall plot, b adjusts signal bandwidth and $N(f,t)$ is Gaussian random field with zero mean and identity covariance matrix. Radon transform recovers the signal as a bright spot with a characteristic hourglass profile. For the plots of n x n pixel resolution Radon integration amplifies the signal by a factor ~n and noise by a factor ~sqrt(n), resulting to typical increase of SNR by a factor ~sqrt(n) relative to the original signal.

## Methods of the analysis

In this paper we will apply Radon transform for the analysis of waterfall plots available at setilive.org. The general scheme of the analysis is shown on Fig.2. The main component, Radon transform is defined by the formula:

$$r(x,a) = \int dy\, s(x \cos a - y \sin a, x \sin a + y \cos a)$$

In numerical estimation the integral has been replaced by a sum over corresponding pixels restricted to the bounds of the waterfall plot. The resulting distribution has been normalized by subtracting the average and dividing to rms. The normalization maps rms→1, so that the resulting 2D plot directly represents Radon transform in SNR units.

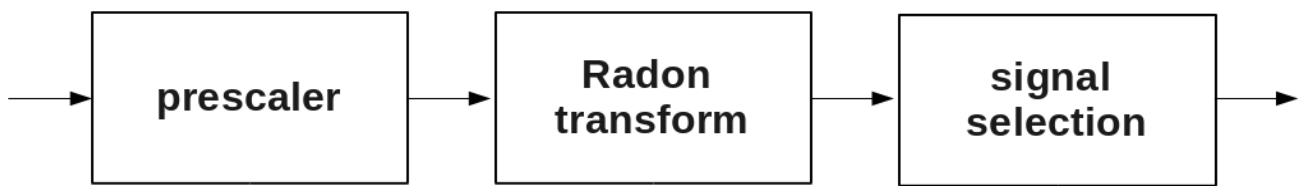

**Fig.2.** Scheme of analysis.

Prescaling step is needed to remap the original waterfall plots to the size appropriate for Radon transform. The original plots are 8-bit grayscale PNG images, coming in two resolutions: 768x384 and 758x410. Since the above described version of Radon transform is suitable for square images, the original plots have been rescaled to 256x256 resolution, using ImageMagick 'convert' tool. In this paper we use direct $O(n^3)$ algorithm for Radon transform. As an improvement, one can replace it with fast $O(n^2 \log n)$ Radon transform and its versions [4-6]. They are designed for the images, whose dimensions are either integer powers of two or prime numbers. The prescaling step will be also necessary for fast transform.

Signal selection is the most important step in the chain. The input waterfall plots come in groups, corresponding to simultaneous signal observations in several directions on celestial sphere. Only the signals highly localized on the sphere are interesting, i.e. the signals present in one observation from the group and absent in others, so called 'single beams'. As the first step, only the groups containing 3 observations have been selected, reducing the total number of groups from 6.73e5 to 2.53e5. Then the signals with small Doppler drift were eliminated. Such signals are produced by the sources which don't move (or have a constant radial speed) relative to the receiver and most probably correspond to terrestrial radio sources or geostationary satellites. These signals are represented as vertical lines on waterfall plots, i.e. as nearly zero angles on Radon plots and can be easily eliminated.

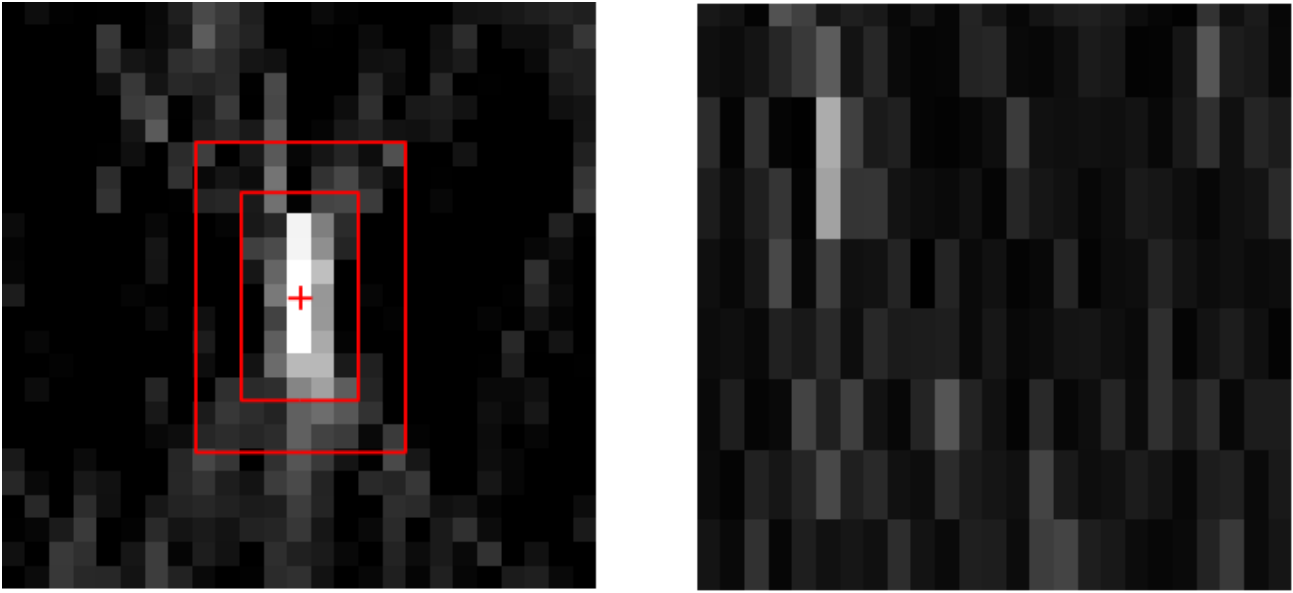

**Fig.3.** On the left: selection of narrow signal on Radon plot.
On the right: vertical correlation on original waterfall plot.

Further, the signals were selected on Radon plots using a method shown on Fig.3 left. Two rectangular contours around a given point were defined. The point is selected if a logical AND of the following conditions took place:

- in the given beam in the given point $SNR_0 > 4$
- in the given beam everywhere between two contours $SNR < SNR_0/2$
- in the other two beams everywhere inside external contour $SNR < 4$

This allows to select narrow signals appearing in a single beam. On Fig.4 left the signals are presented using as coordinates SNR in main beam and maximum of SNR in two other beams. In more detail, in the main beam $SNR_0$ in the given point is taken, while in two other beams the maximum SNR inside the external contour around the given point is taken. The grid step on this plot equals 1. The diagram can be separated to random background, a region of radio frequency interference and a region for potential extraterrestrial signals. The latter region corresponds to strong signal in main beam and weak signals in the other beams. Fig.4 right shows a cumulative distribution plot, where the vertical axis presents the number of selected events possessing SNR in the main beam greater than a value given on the horizontal axis. In appropriate normalization this plot presents an empirical distribution function, a finite sample approximation to cumulative distribution function (CDF), characterizing the probability P(SNR>val) in its tail region.

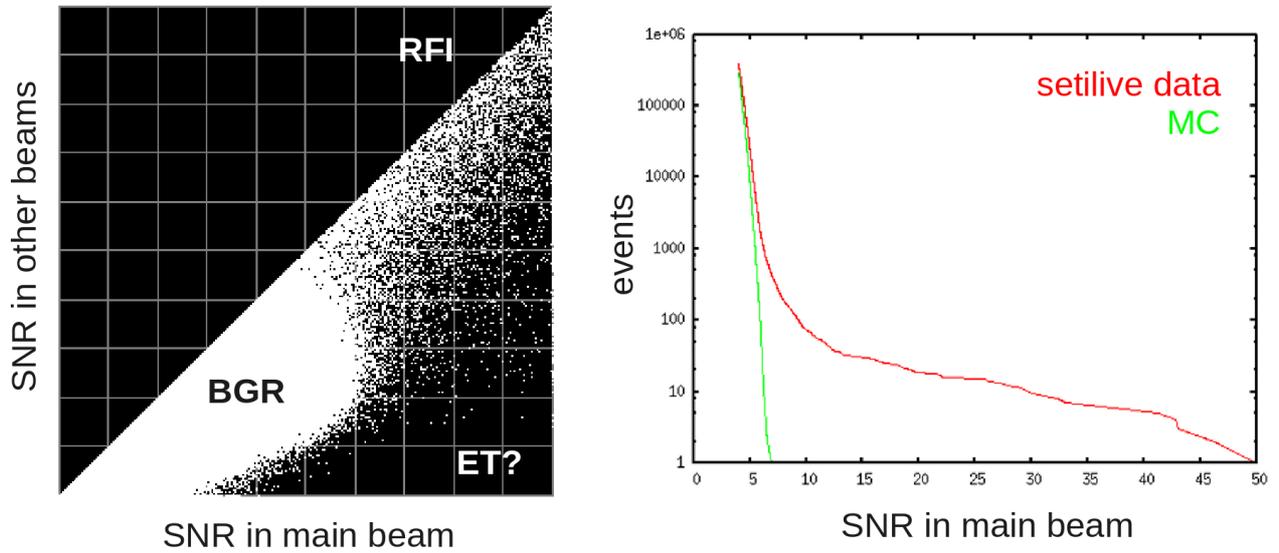

**Fig.4.** On the left: separation of RFI and potential ET signals.
On the right: distribution of selected signals in comparison with Monte Carlo simulation.

**Monte Carlo simulation**

For comparison, we show on Fig.4 the results of Monte Carlo simulation, where Radon transform and the same selection criteria were applied to purely random field. Significant difference of these distributions starts at about SNR~6, which can be used as a threshold separating signals from the background.

In more detail, Radon integral contains lines with different number of pixels e.g. comparing locations in central region and corners of waterfall plots. To compensate this effect, the result of Radon transform before global normalization has been divided to sqrt(k), where k is the actual number of pixels counted along the integration line. For Gaussian random field with uncorrelated pixels this will correct SNR for integration lines of different lengths. However, detailed look into original waterfall plots Fig.3 right shows a presence of vertical correlation. Namely, p consequent pixels in vertical direction share the same value, where p varies between 2-4 and most often equals 3. The reason for this correlation could be in the details of measurement procedure, data processing or final representation of images. We will not speculate, but introduce a correction for this effect. For this purpose we assume the groups of p=3 input vertical pixels strongly correlated, count the number of uncorrelated pixels k' along integration lines and multiply the result of Radon transform to sqrt(k'/k) before global normalization. The same pattern of p=3 vertically correlated random pixels has been also used in MC simulation.

We note also that vertical correlation of Radon plots is not directly related to vertical correlation of waterfall plots. E.g. taking Gaussian random field with identity covariance matrix on input

$$\text{cov}(s(x,y),s(x',y'))=\delta(x-x')\delta(y-y')$$

we obtain the following covariance matrix on output of Radon transform:

$$\text{cov}(r(x,a),r(x',a'))=|\sin(a-a')|^{-1}\chi(x,a;x',a')$$

Here $\chi$ equals 1 if two straight lines defined by (x,a) and (x',a') have intersection inside the domain of random field definition and 0 otherwise. For the random field defined on a square $[-1,1]^2$

$$\chi = \theta(|\sin(a-a')|-|x \sin a'- x' \sin a|) \, \theta(|\sin(a-a')|-|x \cos a'- x' \cos a|)$$

where θ is Heaviside step function. In particular, we have

- cov(r(x,a),r(x',a'))=0 at a=a', x≠x', in horizontal direction of Radon plot;
- cov(r(x,a),r(x',a'))~ $|a-a'|^{-1}$ at a-a'→0, x=x', |x|<1, in vertical direction.

Simulation of random fields with a given covariance matrix can be performed with the methods [7] using principal component analysis (PCA). There are also methods for simulation of random fields with unknown covariance matrix based on numerically efficient singular value decomposition (SVD) of a representative set of samples from the field [8]. These methods can be used as an improvement of MC technique used here for estimation.

In our analysis the processing of real data and MC simulation used the same number of waterfall plots, they were performed on parallel CPUs and were accomplished in the same time. Although such synchronization is not obligatory, since the simulated cumulative distribution could be scaled to the size of real data with an appropriate factor. The results of MC simulation with SNR~6 have a form of bright spots on Radon plots and faint lines on original waterfall plots. They are visually indistinguishable from weak signals we are looking for. The only difference is that for MC results we know definitely that they have random origin. Their small probability was amplified by a large statistics and a narrow filter tuned to the signals of such shape. Although such weak signals cannot be distinguished from random events by their shape, the number of observed signals at SNR~6 should be considered as statistically significant, since it prevails over the estimated number of signals randomly adopting the shape of straight lines.

Thus we leave in selection all signals with Radon's SNR>6. This threshold separates ~1500 events. Our selection algorithm can trigger several events per one signal, because several neighbor pixels of Radon plot can pass selection criteria. In the list below the duplicates will be eliminated, all events originating from the same observation will be represented by only one observation code, comprising 1072 signals in total.

**The results**

The figures below show the brightest signals from the selected set, while the Appendix contains all selected signals. We see that the algorithm works stable also in the presence of several straight lines on the waterfall plot, since they become separable on Radon plot. The lines on the waterfall plots can be reconstructed from selected pixels on Radon plots, actually applying inverse Radon transform to them.  As a further improvement of the algorithm, one can apply the inverse transform not to single pixels but to a group of pixels near selected signal on Radon plot or perform non-linear transforms like contrast corrections r→$r^\gamma$ on Radon plot and apply the inverse Radon transform to the result.

The signals in Appendix are selected 'with a margin' and can require further filtration:

- Many selected signals possess only a slight slope to the vertical on waterfall plots. Their selection depends on the threshold for vertical lines elimination. Currently this threshold has been selected to ±2 pixels in vertical direction from the center of 256x256 Radon plot, corresponding to 2° slope to the vertical on original waterfall plots.
- Many selected signals have a periodical structure combined in parallel lines, e.g. ed9f on Fig.7. This implies the same source for several related frequencies. The signals of this kind can be attributed to terrestrial or satellite RFI. On Radon plane they create a sequence of

- spots located on the same horizontal line. Currently all such signals are preserved in the list. Further the algorithm can be upgraded for automatic recognition of such sequences and their identification as multiple components of the same signal.
- The signals with SNR<7 are almost impossible to recognize on waterfall plots. Their presence is important nevertheless. Firstly, one can see them on Radon plots and this gives enough evidence for their existence. Secondly, the number of reconstructed signals at this level in real data prevails over MC results. These signals should be subjected to further analysis to identify the reason of such excess.

Finally, selected single beams can be subjected to more sophisticated analysis [1], involving the tests for reproducibility of the signal in repeating observations in a given beam, the absence of signal with similar parameters in other beams, not only for simultaneous observations but also in all observations within a certain time frame, etc.

**Conclusion**

We have performed statistical analysis of data stored in SETILive archive, using Radon transform and specially constructed filter for selection of single beams. 28 strong signals and 1072 signals in total were selected at the level SNR>6 in the main beam on Radon plot. This level corresponds to the point where CDF of the signals changes its behavior and starts to differ from the results of Monte Carlo simulation of Gaussian random field.

We will greatly appreciate continuation of SETILive project, which will also make meaningful further developments of the described methods including

- real-time data analysis based on fast Radon transform
- fast inverse Radon transform for reconstruction of selected signals
- parallelization to several processors or/and GPUs
- Monte Carlo simulation of random fields using PCA/SVD
- clustering techniques for detection of multicomponent signals
- distribution of selected signals over observation targets

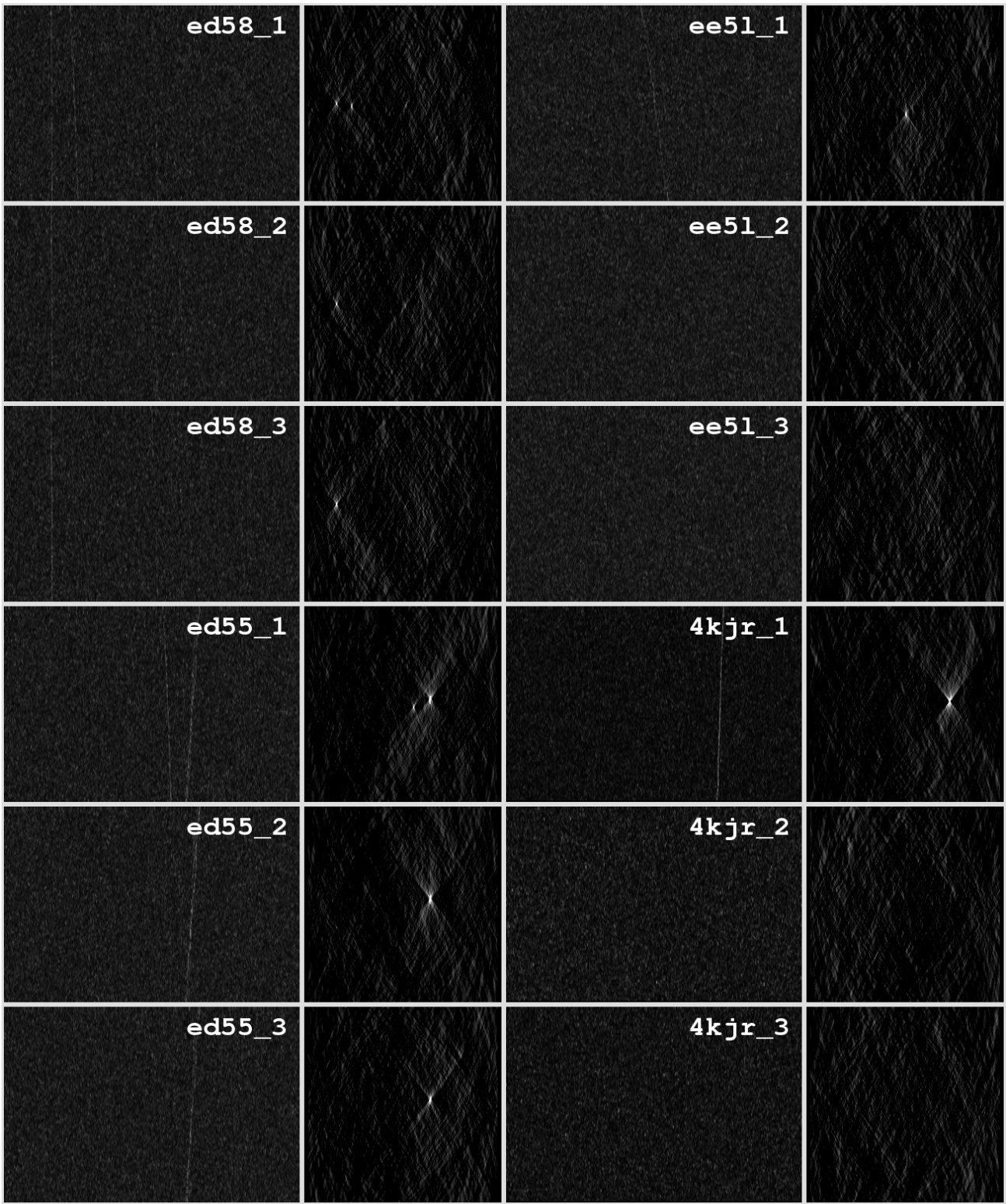

**Fig.5.** Selected signals with highest SNR. The plots for all 3 beams are shown. Radon plot is shown on the right of every waterfall plot.

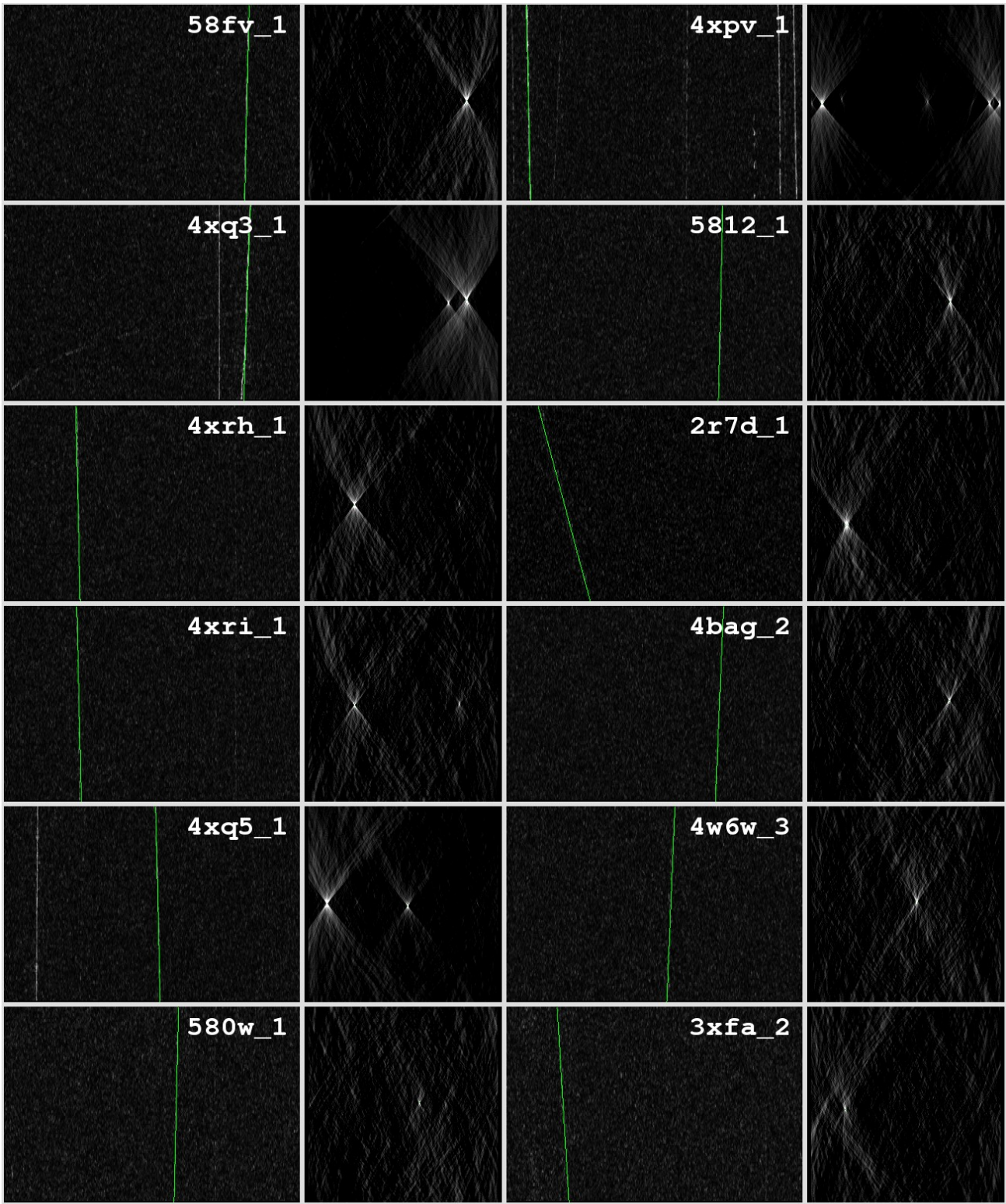

**Fig.6.** Selected signals with highest SNR (cont'd). Reconstructed signals on waterfall plots are shown.

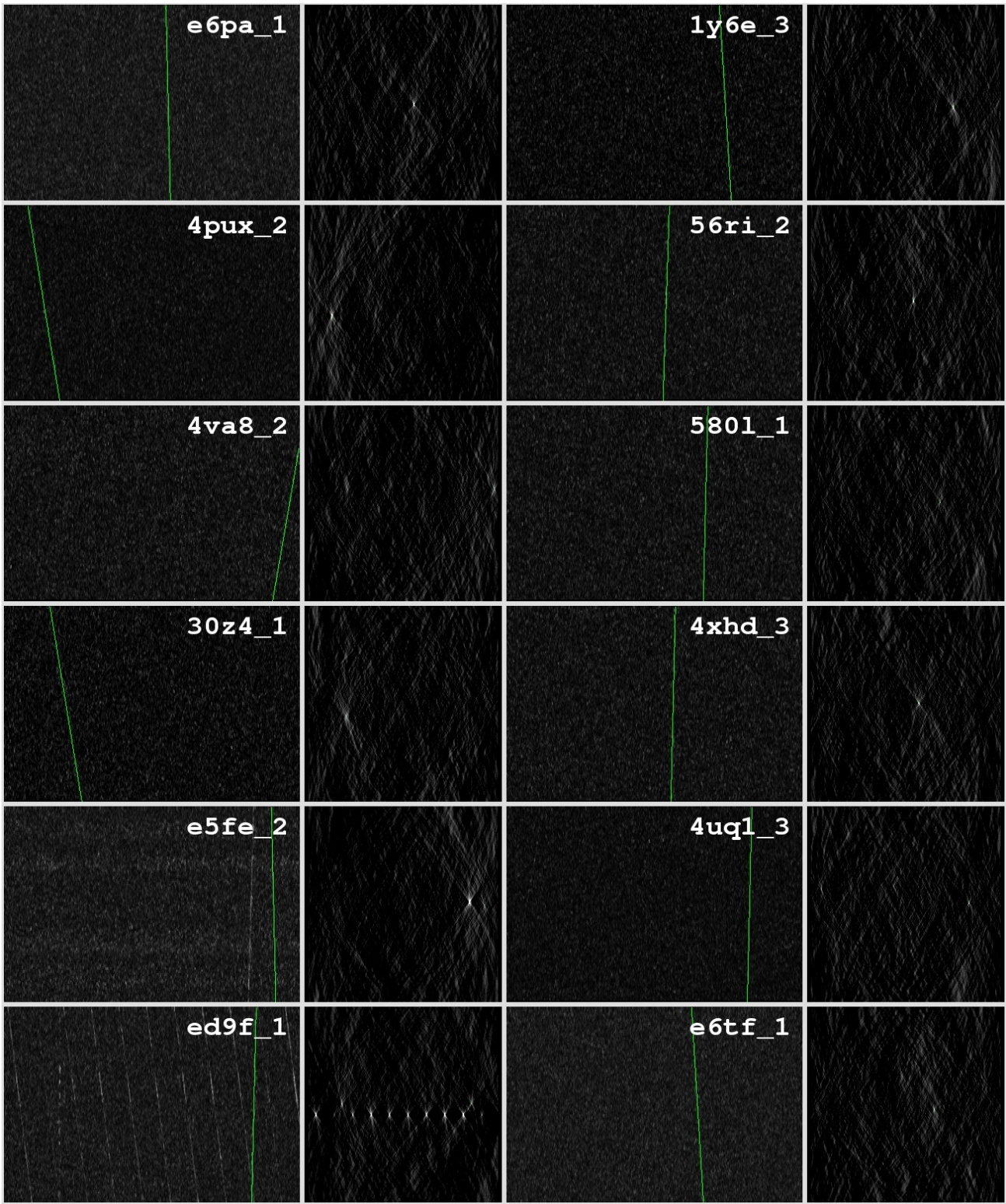

**Fig.7.** Selected signals with highest SNR (cont'd).

**Appendix:** observation codes for all selected signals in SNR descending order

```
58fv_1  4kjr_1  4xpv_1  4xq3_1  5812_1  4xrh_1  2r7d_1  4xri_1  4bag_2  4xq5_1  4w6w_3  580w_1  4sly_1  ed55_1  3zmc_2  ed58_1
3xfa_2  ee5l_1  4qds_3  4x06_2  2m46_1  4xej_2  4tme_3  e6pa_1  1y6e_3  4pux_2  4umi_1  4w4a_3  4wvx_2  4xhj_3  4gke_2  3ylb_3
3zqo_1  4x07_2  56ri_2  4djp_2  4qkj_3  4m59_3  4wt6_2  4qel_3  43y2_2  47wb_2  4va8_2  1x6w_1  4slx_1  47us_3  18jb_1  45c4_3
4uz4_2  3w8g_1  4vkb_3  4xsd_1  41j6_3  4ai4_1  5295_2  580l_1  30z4_1  451y_2  4u5z_3  4q7r_3  3xzq_3  4h65_1  3y8a_3  4hjt_1
4u5x_3  414p_3  45c9_3  4lkx_1  4slu_1  4pzj_3  4xhd_3  4kfo_2  2535_2  40ag_2  4pz4_3  3xzp_3  e5fe_2  4is7_1  3unw_2  2rcm_1
42dz_2  1ift_3  45m0_2  4vcz_2  4ko6_1  4wwg_2  4kg3_2  48gd_1  49m7_2  3yhh_2  4x05_2  47xl_3  4uq1_3  3wh5_1  4l9m_2  4pz1_3
ed54_1  432m_2  3zmt_3  4x0l_3  3xzx_1  3eio_1  4qtb_3  1x70_1  1jkj_3  41re_2  4v3x_1  ed9f_1  45my_1  4bzi_2  3aun_2  4q7p_3
4ypf_2  47t8_3  4kfk_2  4e9b_1  4lld_3  4iqo_2  3ylc_3  3zqv_1  e6tf_1  4v57_3  4aps_3  430y_3  4dp0_2  1cnx_3  4i8m_2  3w5h_1
415u_2  4fob_1  3tpp_2  4ahh_3  3yly_1  4uh6_3  4g2q_2  4jxw_2  4ph9_2  4jq4_1  4jdc_2  4mhi_2  4pqw_3  45bz_3  57ne_1  4avh_3
1img_2  3y7p_2  4wvw_2  42a1_3  4isf_2  4c7i_1  32dl_2  3xzr_1  4tlv_3  agwm_3  3zm8_1  4tar_2  4xcl_1  4grv_3  cc7n_1  0wyn_1
4s8a_2  40ks_1  48ga_2  3vvn_3  418z_3  42ag_1  4ad1_3  40zu_1  45eu_1  4qky_2  48g7_2  40kl_1  ed4u_3  4xc5_2  4kg1_2  2nzv_1
4myw_3  4l11_1  4n2e_3  4wwe_2  edx0_2  415c_2  432s_1  edw6_1  3wt4_2  4fqu_2  ed8d_3  4999_3  4g5s_3  1osp_3  49v8_1  4v8y_2
428d_3  dwlt_2  4myv_3  3xzs_3  4ipq_3  47r1_2  3zm9_1  421g_2  4g2t_1  3zxc_3  4rv4_2  4pvg_2  3uzi_3  4qdr_3  4kfr_2  3zqz_1
40kq_1  42e5_3  4t6c_1  4e9z_2  4eql_3  4ktc_3  4foe_3  4wzz_2  415q_2  4j0w_1  4q5z_2  0tr6_3  4ph9_1  4bwz_1  4wqr_1  4pjk_2
56ud_3  1hxz_1  42sd_1  4ktq_3  4g5z_1  3tpo_2  1uct_1  3zsq_2  3zqr_1  3a9a_2  4i2g_3  4kap_3  432j_2  27nt_1  4c2m_1  3vxq_3
418d_1  2esf_3  2hyz_3  4th2_3  235c_3  4tlu_2  41w7_3  e5mw_2  4dos_2  42n0_3  4jzc_2  4qim_3  2in1_1  2iob_1  2ipo_1  2irq_1
2iz6_1  3w5j_1  dtps_2  42t4_2  4pxo_2  2l3x_2  4ld2_1  35o2_3  4uq6_3  45pf_3  4jt5_2  4i7z_2  3lqr_2  4e4p_1  e6wa_1  e5mw_3
432f_2  42bl_1  1l7d_2  3fhq_2  34w4_2  3aij_3  42gn_1  4327_1  160l_3  c4ht_1  1acx_3  4h4w_2  3vxt_1  1c2u_3  1quz_1  15cy_3
4jdv_2  41ny_3  aduo_3  0u3l_3  28xr_2  4308_3  4grz_3  1bsb_1  4w2u_3  2xmq_3  42vg_3  42ow_1  4pv2_2  c7qh_3  42a2_1  3dbz_3
3gin_1  2v47_3  3w8e_1  3xom_2  1put_1  11g9_3  3jdt_2  31fm_1  44du_2  4ai5_1  0waf_2  3wc1_2  496z_1  415b_3  4nu4_3  agkt_3
4ewz_3  0vnc_2  asax_3  440s_2  45og_1  1iiy_3  45fd_2  4ic0_1  4od0_3  dv5v_1  4bqx_1  249h_1  4zmg_3  as01_2  dua4_3  2zch_3
4e8l_2  4lpb_1  3jji_1  4uxq_3  3y2p_1  asph_1  2ly6_2  456c_1  3rde_2  4tko_3  4l17_3  4328_2  4frp_1  3uzo_1  1y6f_3  1jw5_2
duq1_2  1oqo_1  4ijo_1  4c2b_3  59tz_2  41nv_3  4jue_1  2mh6_2  ajso_1  30wr_3  2anr_2  2cyy_1  24ir_2  4ru8_1  1gbc_1  e0rd_1
4b7t_2  431e_3  1zx4_3  4drx_1  2rvn_3  1dl8_2  421r_2  4suz_3  3q4g_3  4grj_3  3zmu_3  49zr_1  2cvd_2  4si1_3  2uoa_3  300v_2
1i44_1  21u7_3  47zw_3  31nl_3  3e0v_3  ars2_2  3hby_2  4q9k_3  1wz1_3  3xzo_1  4242_1  4t9s_3  bs7k_2  c3ea_1  c1v7_1  24yd_2
ae5d_1  aois_3  10mh_1  4g5r_3  4j6p_3  2t09_1  2o99_3  57nm_2  1gkr_2  2kf5_1  2kfw_1  1t14_1  4306_3  4e99_1  ee1i_1  1wtc_3
50lb_3  2f7v_1  2ihs_1  216d_2  22ub_1  ads0_2  4voa_2  3xzu_1  1lbz_3  2g5f_1  39yu_2  39xh_1  145a_2  3546_3  4hty_3  4h79_3
1lkt_1  3dsb_2  11ce_1  bqlg_2  4e4u_1  42vg_2  4pxh_2  aquk_3  4ugz_2  2obl_2  2oqg_2  1hxv_1  3zmi_1  3tim_2  21ny_1  4c80_1
1qkn_3  286w_2  1jeh_2  3rq4_2  4u6g_3  2y9t_2  2wu1_3  1hq2_3  du2d_3  196r_3  3uwn_1  e3e1_2  30wo_1  3zsr_1  4ade_3  0y58_1
4mls_3  195m_3  198n_3  0szo_3  4u9y_3  1nss_3  10wp_3  1l61_2  126g_1  511n_3  4ejp_1  4dsa_1  421q_1  3wkt_2  1doq_3  4xfb_3
aq2o_2  2y4w_3  1re0_2  0t0l_2  4keg_1  2rys_3  523x_1  4fyq_1  4jte_3  4tr9_3  4iro_1  29bl_1  1psq_1  29ha_1  3y02_3  4e4b_2
c8ef_3  2lws_1  4mde_1  5c6y_2  2lqq_2  42cj_3  5b18_3  30x5_3  49m9_3  4n2e_1  aflm_1  4183_2  3v2q_3  4s0d_1  dz2f_1  2gny_1
1n1q_1  2kwm_2  5919_1  dxw9_2  dnkq_1  1710_1  du6x_1  5bl6_1  3a48_2  4itj_1  2rd0_1  3v6t_1  1yvu_2  44ag_3  28uw_3  4rk8_2
c2tf_1  0us4_3  125x_3  29y1_1  176p_3  40mr_2  28tn_3  arsb_1  aqy6_2  1ftg_3  3ut0_1  atip_3  2h0l_1  45pz_2  atfr_1  3wux_3
c9ga_3  5037_1  eds8_3  3wkl_1  4tf2_3  2no2_3  1b8x_1  1kcj_3  2w0h_3  1u6p_1  42t2_3  45t9_2  321j_1  4e20_3  4974_1  0zk6_3
1ly6_3  4hjs_1  bpz6_3  1s4t_2  0u17_1  42eg_2  4mlp_3  ccfj_2  4xhn_3  2m43_1  bpt4_3  4973_1  432j_1  1jan_3  3uwl_2  29z8_3
4twe_1  2csc_3  2x4j_2  dva2_1  3sp6_1  3s9y_2  cak6_2  426h_1  4v4f_1  4lcp_2  43vx_3  4rhm_1  4uxs_3  4xlb_2  2cgr_3  42eb_3
36r3_3  5dax_1  aehz_1  edkx_3  4tgr_2  3y18_2  dysq_2  502c_3  2h31_3  3flh_2  2rtw_2  1t3v_1  ajue_1  4bls_1  37gt_1  akzx_2
4vl3_3  42b2_2  5b2c_1  2yd1_3  0t7u_2  1u2y_1  3hov_2  2wm5_3  1ulc_2  e1qr_2  2dl4_2  aehp_1  3wes_1  57h5_1  5e7c_2  4lf0_1
cauw_1  4x83_3  3w8f_1  4hi9_2  1aqq_3  43c1_3  426j_1  1o8e_2  e329_2  3trs_3  2to9_3  1b4j_3  4jdd_2  1acj_1  0sw7_3  31ol_1
14yn_3  4ea0_2  dyln_2  4e9x_2  51xs_3  4qt3_1  40kk_1  4rtz_2  5851_3  c5lu_3  4j5i_1  4w85_2  4rid_3  5eur_2  e20e_2  4hwb_2
31g0_1  3ajq_3  33er_2  4x01_2  1q6d_2  2fho_1  3zxb_3  499f_3  1q5f_3  4e4c_2  10dd_2  ate2_1  4k7q_2  2yle_1  2nrv_1  294b_2
58r5_2  4sv4_3  3c5k_3  0trj_1  2z7h_3  201o_1  dk27_2  3uwx_1  bn10_2  31xr_2  cevi_2  cdyc_3  0zx5_3  2ljz_3  4l3c_1  2xk3_2
0ukb_2  2rra_3  4w9g_3  2tfc_2  agl5_3  as8a_2  e4gc_3  39gm_2  2brp_2  1g8m_2  34qu_1  4hni_2  e8cp_3  2krg_3  0wrh_1  37bb_3
5a3w_2  1adv_2  3xq3_2  3jxk_3  1s6u_2  2w5c_3  dv8e_1  5d0h_1  3021_1  22pb_1  4e9w_1  18yo_2  3c7i_2  424e_1  c8f1_2  4c0f_3
2518_2  3wjj_2  2pm1_3  29jc_2  2kdl_2  4w2y_2  4e25_1  atux_2  5exy_1  3g65_1  0y7i_3  1ig7_2  4jq1_1  2d7g_1  bw0r_3  alcb_3
e172_3  2dy3_3  4doz_3  am57_2  2yds_1  2cof_3  5cer_3  3pjy_1  c78t_3  4u5a_1  3h5u_1  3zqq_3  anad_3  48wr_3  1tbh_1  51e4_3
0z2a_1  ee5d_2  41it_2  27np_3  ameg_1  3hlc_3  1wio_2  4iy2_1  1c9q_2  57jg_1  25ae_3  49i0_1  442j_3  271j_1  13ox_2  2wc0_2
39gi_1  4x1a_3  1vtw_3  2giv_2  2d0g_1  41t7_3  edw9_1  2a71_3  48tf_1  438q_1  1win_2  354z_1  45pb_3  dzf5_2  aizc_3  1aqh_1
boti_2  39ql_3  1uqe_1  0zzr_3  1fpa_3  36iv_3  3rsc_1  2g2d_3  1w2t_2  1aym_1  2yt5_3  4mrc_2  4i5n_2  57kr_1  dxpt_1  4jlz_2
c68f_1  4oid_2  c17b_1  ceer_3  1xwz_1  28fo_2  28zo_2  5a41_1  djzd_1  32kz_1  2cc2_3  bpw9_1  1ns0_1  4y0f_1  430v_3  4bqx_2
2lw7_3  dmui_3  2pfw_1  0un2_3  c4rk_3  3thk_3  3ti2_3  4bgn_3  1qea_3  3nrn_2  45c1_1  2527_3  58eu_1  42c0_3  2qyt_2  58vk_1
2yya_2  40c0_3  5ct3_2  1kch_1  1wti_3  582v_1  c6hk_1  aft5_2  23ni_1  22y4_1  4hjp_1  0z8q_3  4ke7_2  5erl_1  4api_3  2xuj_1
4zaw_1  dyiw_2  artw_2  ci34_3  2orf_1  149h_1  3onc_2  3934_3  4yg0_1  1d4r_2  3bcn_2  27h5_2  adb2_2  5fce_1  5de6_2  23gc_3
2buy_3  1gyh_3  ajk3_2  2zmu_3  56ra_2  bxwc_2  4zm8_2  1xsz_2  2jbj_1  4ke8_1  26mf_3  15z9_3  cgir_1  3occ_2  3ny5_2  0uaa_3
42aw_2  32jt_1  2kjv_1  47r4_3  3kw7_3  21gb_1  2ua2_2  bylv_2  1s77_2  1ofs_2  3587_3  dlwc_3  2y7i_2  1ok0_2  agje_2  afke_1
4c5s_1  brdu_1  2g56_2  acpi_1  3yhi_2  4306_2  4bbp_3  4zls_3  2o7r_2  bn8c_1  4ew6_3  239d_3  51tt_3  c7nn_3  2vwr_1  44ff_1
0ukn_1  3m53_1  3vvl_1  0zsc_1  3f89_3  4ry2_3  4jq5_1  1ln4_1  laor_3  4x35_2  2cul_1  3dc7_2  2rtr_0  2tr6_3  4wk2_2  4318_3
2jff_2  3smd_1  1ipp_3  2u6h_3  36yq_2  4huj_3  2c5v_2  4u6e_3  4ka8_3  5c1w_1  0xvc_2  1wlc_3  3y9t_2  40f3_3  2r91_2  at0c_2
e8gb_2  ccrm_1  30ev_1  0ywu_1  4prx_3  dvm6_2  4ktp_3  2blk_3  29lb_1  3oa4_2  441s_1  4dkv_3  4ujv_1  2y9n_1  2a5z_1  3zyy_1
4r5e_2  4479_2  0wzb_3  dvf6_3  4qsi_1  2o1l_3  0y12_3  ci3m_3  1jha_3  0xef_3  0sym_2  1wn8_2  45iw_3  0ts9_2  1urc_1  ac4o_2
18hi_1  3uaj_1  57on_1  4aj0_1  3e13_1  akx8_2  56uk_3  2jhu_2  2zbe_3  42c0_2  1wup_2  4nk3_2  4xa6_3  2dss_3  4et6_1  13uv_2
4ke9_1  42a0_2  28hi_3  cagh_3  121i_3  2ma6_1  1xmr_3  4ijo_3  3pb8_3  38yw_3  4l9v_3  35sz_1  anhe_1  4qp2_3  58lg_1  4phz_1
38zp_3  0szr_1  1cfn_2  3rwg_1  2t2j_2  2x7v_2  1xad_1  2v52_1  4kfu_2  4ll6_3  49xy_3  dm17_1  17ca_2  dliu_1  432a_2  3uss_1
ae56_3  4yra_1  1ibu_1  2ik3_1  2ike_1  2ilg_1  2inv_1  2iq3_1  3jlb_3  0x5s_3  1ugv_3  bnmh_3  bnaz_2  3kh7_1  3mye_1  ciq6_3
1qfi_2  11x3_2  577c_3  4n9h_1  3grr_2  edyk_2  29k5_2  3qyt_1  3xwd_1  1h5g_2  5c9t_3  4y9k_1  c51z_1  bq3k_2  48cz_1  cemx_1
2l11_2  5awe_3  2qp4_1  1btl_3  dxna_3  4u2h_1  22zg_1  4eri_3  4j5j_3  dziq_2  adub_1  4g1i_3  4sth_3  40zz_2  cgay_2  1d8p_1
```

Here the last digit defines beam number and the preceding 4-character code can be used to retrieve the waterfall plots from

http://talk.setilive.org/observation_groups/GSL000****